\documentclass[12pt,harvard]{iopart}

\usepackage{iopams}

\usepackage{graphicx}
\usepackage{natbib}
\usepackage{amssymb}
\usepackage{graphicx}
\usepackage{subfigure}

\newcommand{\be}[1]{\begin{equation}\label{#1}}
\newcommand{\ee}{\end{equation}}
\newcommand{\ba}[1]{\begin{eqnarray}\label{#1}}
\newcommand{\ea}{\end{eqnarray}}
\newcommand{\rf}[1]{(\ref{#1})}
\newcommand{\nn}{\nonumber}
\newcommand{\diag}{\mbox{\rm diag}\,}

\begin{document}

\title[]{Instabilities of rotational flows in azimuthal magnetic
fields of arbitrary radial dependence}

\author{Oleg N. Kirillov$^1$\footnote{Corresponding
author: o.kirillov@hzdr.de}, Frank Stefani$^1$ and
Yasuhide Fukumoto$^2$}

\address{$^1$Helmholtz-Zentrum Dresden-Rossendorf, P.O. Box 510119, D-01314 Dresden, Germany\\
$^2$Institute of Mathematics for Industry, University of Kyushu, 744 Motooka, Nishi-ku, Fukuoka 819-0395, Japan}

\ead{o.kirillov@hzdr.de, f.stefani@hzdr.de, yasuhide@imi.kyushu-u.ac.jp}

\begin{abstract}
Using the WKB approximation we perform a linear stability
analysis for a rotational flow
of a viscous and electrically conducting fluid in an external
azimuthal magnetic field that has an arbitrary radial profile $B_{\phi}(R)$.
In the inductionless approximation, we find the growth rate of the
three-dimensional perturbation in a closed form and demonstrate
in particular
that it can be positive when the velocity profile is Keplerian and
the magnetic field profile is slightly shallower than $R^{-1}$.
\end{abstract}

\vspace{2pc}

\maketitle

\section{Introduction}
Recently, the magnetorotational instability (MRI), discovered by
Velikhov (1959) and Chandrasekhar (1960), has
experienced a significant revival for its promise to explain
the destabilization and turbulization of Keplerian disks and,
consequently, the outward transport
of angular momentum and the accretion of matter to a gravitating central
body such as a black hole or a protostar (Balbus 2003).
Complementary to the extensive numerical research with main
focus on astrophysical applications,
there is also some activity to reproduce the MRI in the liquid metal
laboratory. Up to present, any clear identification of the
standard version of MRI (SMRI), with an axial magnetic field being applied,
has been hampered by the necessity to reach quite large magnetic
Reynolds numbers (in the order of 10). Since liquid metals are characterized
by a very small magnetic Prandtl number $\rm {Pm}$ (the ratio of viscosity to magnetic
diffusivity), this implies that the hydrodynamic Reynolds numbers must be in the order
of millions, a number at which experimental Taylor-Couette flows are very hard to
maintain stable due to the effects of axial boundaries.

It came, therefore, as a great surprise when Hollerbach and R\"udiger (2005)
discovered an alternative version of the MRI whose onset does not depend on the magnetic
Reynolds number and the Lundquist number, such as SMRI, but only on the Reynolds number
and the Hartmann number. This new MRI version relies on an appropriate combination of an
axial and an azimuthal magnetic field, and has therefore been coined helical
MRI (HMRI). Further, a non-axisymmetric version, the azimuthal MRI (AMRI),
has been shown to become dominant for purely or strongly dominant azimuthal magnetic
fields (Hollerbach et al. 2010, R\"udiger et al. 2013).

Whereas the scaling of HMRI (and AMRI) with the Reynolds and Hartmann numbers
is quite attractive for experimental studies (Stefani et al. 2006, 2009),
a somewhat unattractive feature of these inductionless MRI versions had been
identified by Liu et al. (2006). It concerns the apparent
failure of HMRI to work for comparably shallow rotation profiles, including the
Keplerian profile that is the most relevant one for astrophysics.
For the inductionless case, with $\rm {Pm}=0$, the authors had identified two
limits for the steepness of the rotation profile, one for negative, the other
one for positive Rossby number $\rm {Ro}$, between which HMRI ceases to exist.
Later, the significance of both limits was extended to the case of AMRI, and
even to higher azimuthal wavenumbers (Kirillov and Stefani 2010, Kirillov et al. 2012).

The relevance of the inductionless limit is not restricted to the academic case of
liquid metal experiments. Small magnetic Prandtl numbers
appear as well in the outer parts of accretion disks around black holes
(Balbus and Henri 2008), in protoplanetary disks (Armitage 2011), as well as in
the liquid metal cores of Earth-like planets (Petitdemange et al. 2008).
Given the dramatically different scaling laws of SMRI and
HMRI/AMRI it is of great importance to determine the
range of applicability of the latter.

In a recent letter (Kirillov and Stefani 2013) we have discussed
a simple way of extending this range. We set out from
the consideration that the current-free assumption for the
azimuthal magnetic field, i.e $B_{\phi}(R) \propto 1/R$, as it
was assumed theoretically and imposed experimentally up to present,
has no particular astrophysical foundation. Quite in contrast,
it is rather clear that $B_{\phi}$ in the disk
may have a complicated spatial structure that is a function of
the axial and radial dependencies of conductivity
and viscosity. This applies both to the case that an
axial magnetic field $B_z$ is externally applied, from which
$B_{\phi}$ is then induced, as well as
to the case $B_z=0$ so that $B_{\phi}$ must be produced by
some sort of dynamo (Herault et al. 2011).
An instructive illustration of the induction of a
toroidal field from an applied poloidal field can be
found in a recent paper by Colgate et al. (2011)
who had observed a 8-fold toroidal field gain
in a liquid sodium Taylor-Couette flow with a
magnetic Reynolds number ({\rm Rm}) of 120. While
certainly this mechanism
cannot be transferred one-to-one to an accretion disk,
which has no Ekman layers at its vertical edges,
there are other vertical and radial variations of angular velocity and
conductivity that
would lead to similar induction effects.
Without going into the details of magnetic field
induction in various types of accretion disks, the
idealized case of a current-free field, $B_{\phi}(R) \propto 1/R$,
as produced by an
assumed  central axial current, is by far less likely than
any shallower profile that is (at least partly) induced in the disk.
Defining an appropriate magnetic Rossby
number $\rm {Rb}$ for such general profiles $B_{\phi}(R)$, we showed that
Keplerian profiles can easily be destabilized by HMRI or AMRI. We further
showed that the upper and lower Liu limits in terms of $\rm {Ro}$ (Liu et al.
2006) are just the endpoints of a continuous stability curve in the
$\rm{Rb}-\rm{Ro}$ plane.

Having thus evidenced that {\it formally} the inductionless versions of MRI,
i.e. HMRI and AMRI, are well capable of destabilizing Keplerian, and even shallower,
rotation profiles, one should notice a physical inconsistency of this
argumentation. Certainly, any
deviation of $B_{\phi}(R)$ from the current-free $1/R$ profile
can only result from induction effects in the disk. Hence it would
need some finite value of $\rm {Rm}$, which is apparently in contrast
with the original inductionless approximation.

As a follow-up of the letter (Kirillov and Stefani 2013),
the present paper comprises  a detailed derivation of the
WKB stability analysis for
arbitrary radial profiles of $B_{\phi}$,
quite in analogy to the procedure by Krueger et al. (1966), Eckhardt and Yao (1995),
and Friedlander and Vishik (1995). A further focus will then be on
some strict results concerning the growth rate of the non-axisymmetric
instability. Finally, using the Bilharz criterion we obtain the domain of AMRI of a Keplerian flow both in the inductionless case corresponding to the vanishing magnetic Reynolds number and in the case of finite $\rm Rm$.

\section{Mathematical setting}
\subsection{Basic equations and the base state}
The standard set of non-linear equations of
dissipative incompressible magnetohydrodynamics
consists of the Navier-Stokes equation for
the fluid velocity $\textbf{u}$ and the induction equation for the magnetic field
$\textbf{B}$,
\ba{m1}
\frac{\partial \textbf{u}}{\partial t}+\textbf{u} \cdot \nabla \textbf{u}-\frac{1}{\mu_0 \rho}\textbf{B}\cdot \nabla\textbf{B} +\frac{1}{\rho} \nabla P-\nu \nabla^2 \textbf{u}&=&0,\nn \\
\frac{\partial \textbf{B}}{\partial t}+\textbf{u} \cdot \nabla \textbf{B} - \textbf{B} \cdot \nabla \textbf{u}- \eta \nabla^2 \textbf{B}&=&0,
\ea
where $P=p+\frac{\textbf{B}^2}{2\mu_0}$ is the total pressure, $p$ is the hydrodynamic pressure, $\rho=const$
the density, $\nu=const$ the kinematic
viscosity,
$\eta=(\mu_0 \sigma)^{-1}$ the magnetic diffusivity, $\sigma=const
$ the conductivity of the fluid,
and $\mu_0$
the magnetic permeability of free space.
Additionally, the mass continuity equation for incompressible flows
and the solenoidal condition for the magnetic induction yield
\be{m3}
\nabla \cdot \textbf{u} = 0,\quad  \nabla \cdot \textbf{B}=0.
\ee

Introducing cylindrical coordinates $(R, \phi, z)$, we consider
the stability of a steady-state background liquid flow characterized by
the angular velocity profile $\Omega(R)$ exposed to an azimuthal
background magnetic field:
\be{m4} \textbf{u}_0(R)=R\,\Omega(R)\,\textbf{e}_{\phi},\quad p=p_0(R),
\quad \textbf{B}_0(R)=B_{\phi}^0(R)\textbf{e}_{\phi}.
\ee
Note that if the azimuthal component is produced by a central
axial current $I$, then
\be{m4a}
B_{\phi}^0(R)=\frac{\mu_0 I}{2 \pi R}.                                   
\ee
In case that it is produced by a homogeneous current
in a cylindrical column of radius $R_0$, which is relevant for the
onset of the Tayler instability (Seilmayer et al. 2012), it would read as follows:
\be{m4tayler}
B_{\phi}^0(R)=\frac{\mu_0 I R}{2 \pi R_0^2}.                                   
\ee

The centrifugal acceleration of the
background flow \rf{m4} is compensated by the pressure gradient
 \be{m4d} R\Omega^2=\frac{1}{\rho}\frac{\partial
p_0}{\partial R}. \ee

An appropriate quantitative measure of the hydrodynamic shear is
given by the \emph{hydrodynamic Rossby number} $({\rm Ro})$
which we define by means of the relation
\be{hro}
{\rm Ro}=\frac{R}{2\Omega}\frac{\partial \Omega}{\partial R}.
\ee
With this definition, the solid body rotation corresponding
to $\Omega(R)=const$ gives ${\rm Ro}=0$, the Keplerian rotation with
$\Omega(R)\propto R^{-3/2}$ gives ${\rm Ro}=-3/4$,
whereas the velocity profile $\Omega(R)\propto R^{-2}$
leads to ${\rm Ro}=-1$.

Similarly, we introduce the \emph{magnetic Rossby number} ({\rm Rb}) as
\be{mro}
{\rm Rb}=\frac{R}{2(B_{\phi}^0/R)}\frac{\partial (B_{\phi}^0/R)}{\partial R}.
\ee
Hence, ${\rm Rb}=0$ corresponds to the linear dependence of the magnetic
field on the radius, $B_{\phi}^0(R)\propto R$, and
${\rm Rb}=-1$ to the radial dependence given by Eq.~\rf{m4a}.

\subsection{Linearization with respect to non-axisymmetric perturbations}

To describe natural oscillations in the neighborhood of the magnetized
rotational flow we linearize equations \rf{m1} subject to the
constraints \rf{m3} in the vicinity of the stationary solution \rf{m4}
assuming general perturbations $\textbf{ u}=\textbf{u}_0+\textbf{u}'$,
$p=p_0+p'$, and $\textbf{B}=\textbf{B}_0+\textbf{B}'$ and
leaving only the terms of first order with respect to the primed quantities:
\ba{L1}
\partial_t\textbf{u}'&+&\textbf{u}_0 \cdot \nabla\textbf{u}'+\textbf{u}' \cdot \nabla\textbf{u}_0
-\frac{1}{\rho\mu_0}\left(\textbf{B}_0 \cdot \nabla\textbf{B}'{+}\textbf{B}' \cdot \nabla\textbf{B}_0 \right)=\nn\\
&+&\nu \nabla^2 \textbf{u}'-\frac{1}{\rho}\nabla p'-
\frac{1}{\rho \mu_0}\nabla(\textbf{B}_0 \cdot \textbf{B}'),\nn\\
\partial_t \textbf{B}'&+&\textbf{u}_0 \cdot \nabla\textbf{B}'+\textbf{u}' \cdot \nabla\textbf{B}_0-
\textbf{B}_0\cdot \nabla\textbf{u}'-\textbf{B}'\cdot \nabla\textbf{u}_0=\eta \nabla^2 \textbf{B}'.
\ea
Here, the perturbations fulfill the constraints
\be{L2}
\nabla \cdot \textbf{u}' = 0,\quad  \nabla \cdot \textbf{B}'=0.
\ee

With the gradients of the background fields represented by the $3 \times 3$ matrices
\ba{L3a}
\mathcal{U}(R)&:=&\nabla \textbf{u}_0=\Omega\left(
                                \begin{array}{ccc}
                                  0 & -1 & 0 \\
                                  1+2{\rm Ro} & 0 & 0 \\
                                  0 & 0 & 0 \\
                                \end{array}
                              \right), \nn\\
\mathcal{B}(R)&:=&\nabla \textbf{B}_0=\frac{B_{\phi}^0}{R}\left(
 \begin{array}{ccc}
                                                                              0 & -1 & 0 \\
                                                                              1+2{\rm Rb} & 0 & 0 \\
                                                                              0 & 0 & 0 \\
                                                                            \end{array}
                                                                          \right),
\ea
the linearized equations of motion take the form
\ba{L5}
(\partial_t+\mathcal{U}+\textbf{u}_0 \cdot \nabla)\textbf{u}'
&+&\frac{1}{\rho}\nabla p'+
\frac{1}{\rho \mu_0} \textbf{B}_0\times(\nabla\times \textbf{B}')\nn\\
&+&
\frac{1}{\rho \mu_0} \textbf{B}'\times(\nabla\times \textbf{B}_0)= \nu \nabla^2 \textbf{u}',\nn\\
(\partial_t -\mathcal{U}+\textbf{u}_0 \cdot \nabla)\textbf{B}'&+&(\mathcal{B}-
\textbf{B}_0\cdot \nabla)\textbf{u}'=\eta \nabla^2 \textbf{B}'.
\ea

\section{Geometrical optics equations}

We seek for solutions of the linearized equations \rf{L5} in the form
of the geometrical optics (or WKB) approximation:
\ba{g1}
&\textbf{u}'(\textbf{x},t,\epsilon)=e^{i\Phi(\textbf{x},t)/\epsilon}\left(\textbf{u}^{(0)}(\textbf{x},t)+\epsilon\textbf{u}^{(1)}(\textbf{x},t) \right)+\epsilon \textbf{u}^{r}(\textbf{x},t),&\nn \\
&\textbf{B}'(\textbf{x},t,\epsilon)=e^{i\Phi(\textbf{x},t)/\epsilon}\left(\textbf{B}^{(0)}(\textbf{x},t)+\epsilon\textbf{B}^{(1)}(\textbf{x},t) \right)+\epsilon \textbf{B}^{r}(\textbf{x},t),&\nn \\
&p'(\textbf{x},t,\epsilon)=e^{i\Phi(\textbf{x},t)/\epsilon}\left(p^{(0)}(\textbf{x},t)+\epsilon p^{(1)}(\textbf{x},t) \right)+\epsilon p^{r}(\textbf{x},t),&
\ea
where $\textbf{x}$ is a vector of coordinates,
$0<\epsilon\ll 1$ is a small parameter, $\Phi$ is
a real-valued scalar function that represents the phase of
oscillations, $\textbf{u}^{(j)}$, $\textbf{B}^{(j)}$,
and $p^{(j)}$, $j=0,1,r$ are complex-valued amplitudes,
see e.g. Eckhardt and Yao (1995) and Friedlander and
Vishik (1995).

Following Landman and Saffman (1987), Dobrokhotov and Shafarevich (1992),
and Eckhardt and Yao (1995) we assume further in the text
that $\nu=\epsilon^2\tilde \nu$ and $\eta=\epsilon^2\tilde \eta$ and introduce the derivative
along the fluid stream lines:
\be{g2}
\frac{D}{Dt}:=\partial_t + \textbf{u}_0 \cdot \nabla.
\ee

Substituting the expansions \rf{g1} into equations \rf{L5},
collecting terms at $\epsilon^{-1}$ and $\epsilon^0$, and eliminating the
pressure by standard manipulations, we obtain the phase equation
\be{g12}
\frac{D\textbf{k}}{D t}=-\mathcal{U}^T\textbf{k},
\ee
where  $\textbf{k}=\nabla \Phi$, and the amplitude (or transport) equations
\ba{g13}
\frac{D\textbf{u}^{(0)}}{Dt}&=&-\left(\mathcal{I}-
2\frac{\textbf{k}\textbf{k}^T}{|\textbf{k}|^2} \right)\mathcal{U} \textbf{u}^{(0)}-\tilde \nu
|\textbf{k}|^2\textbf{ u}^{(0)}\nn\\
&+&
\frac{1}{\rho\mu_0}\left(\mathcal{I}-\frac{\textbf{k}\textbf{k}^T}{|\textbf{k}|^2}\right)
\left(\mathcal{B}+\textbf{B}_0\cdot \nabla\right)\textbf{B}^{(0)}, \nn\\
\frac{D \textbf{B}^{(0)}}{Dt}&=& \mathcal{U} \textbf{B}^{(0)}-\tilde \eta|\textbf{k}|^2\textbf{B}^{(0)}-
(\mathcal{B}-\textbf{B}_0\cdot \nabla) \textbf{u}^{(0)},
\ea
where $\mathcal{I}$ is a $3\times 3$ identity matrix.
In the absence of the magnetic field these equations
reduce to those obtained by Landman and Saffman (1987), Dobrokhotov and Shafarevich
(1992), and Eckhardt and Yao (1995).

\subsection{Amplitude equations}

Let the orthogonal unit vectors ${\bf e}_R(t)$, ${\bf e}_{\phi}(t)$, and
${\bf e}_z(t)$ form a basis in a coordinate system moving along the
fluid trajectory.
With ${\bf k}(t)=k_R {\bf e}_R(t)+k_{\phi} {\bf e}_{\phi}(t)+k_z {\bf e}_z(t) $,
${\bf u}(t)=u_R {\bf e}_R(t)+u_{\phi} {\bf e}_{\phi}(t)+u_z {\bf e}_z(t) $,
and with the matrix $\mathcal{U}$ from \rf{L3a}, we find that
\be{ae1}
\dot{\bf e}_R=\Omega(R){\bf e}_{\phi}, \quad \dot{\bf e}_{\phi}=-\Omega(R){\bf e}_R.
\ee
Hence, the equation \rf{g12} in the coordinate form
$$
\dot k_R-\Omega k_{\phi}=-\Omega k_{\phi} -R \partial_R
\Omega k_{\phi}, \quad \dot k_{\phi}+\Omega k_R=\Omega k_R, \quad \dot k_z=0
$$
yields
\be{ae2}
\dot k_R= -R \partial_R \Omega k_{\phi}, \quad \dot k_{\phi}=0, \quad \dot k_z=0.
\ee

According to Eckhardt and Yao (1995) and Friedlander and Vishik (1995),
in order to study physically relevant and potentially unstable modes we
have to choose bounded and asymptotically non-decaying solutions of
the system \rf{ae2}. These correspond to $k_{\phi}\equiv 0$ and $k_R$ and $k_z$ being
time-independent.

Denoting $\alpha=k_z|\boldsymbol{ k}|^{-1}$, where $|{\bf k} |^2=k_R^2+k_z^2$,
we find that $k_R k_z^{-1}=\sqrt{1-\alpha^2}\alpha^{-1}$ and write the
partial differential equations \rf{g13} for the amplitudes in the
coordinate representation.
Assuming the solution to the resulting equations in the modal form
$e^{\gamma t +im\phi+i k_z z}$ (Friedlander and Vishik 1995), and
taking into account that $B_R^{(0)} k_R+B_z^{(0)}k_z=0$ in the short-wavelength
approximation, we single out the equations for the radial and azimuthal
components of the fluid velocity and magnetic field. Introducing the
viscous and resistive frequencies as well as the $\rm Alfv\acute{e}n$
frequency of the azimuthal magnetic field
\be{freq}
\omega_{\nu}=\widetilde{\nu} |\boldsymbol{ k}|^2,\quad \omega_{\eta}=\widetilde{\eta}|\boldsymbol{ k}|^2,\quad  \omega_{A_{\phi}}=\frac{B_{\phi}^0}{R\sqrt{\rho\mu_0}},
\ee
so that, in particular,
\be{alf}
{\rm Rb}=\frac{R}{2\omega_{A_{\phi}}}\frac{\partial \omega_{A_{\phi}} }{\partial R},
\ee
we finally obtain the amplitude equations as follows
\ba{mri4ac}
(\gamma +  im \Omega  +\omega_{\nu}) u_R^{(0)}   &-&2\alpha^2\Omega u_{\phi}^{(0)} +2\alpha^2\frac{\omega_{A_{\phi}}}{\sqrt{\rho \mu_0} }B_{\phi}^{(0)}-\frac{im \omega_{A_{\phi}}B_R^{(0)}}{\sqrt{\rho \mu_0}}=0,\nn\\
(\gamma +im \Omega +\omega_{\nu}) u_{\phi}^{(0)}  &+&2\Omega(1+ {\rm Ro})u_R^{(0)}-\frac{2\omega_{A_{\phi}}}{\sqrt{\rho \mu_0}}(1+{\rm Rb})B_R^{(0)}\nn\\&-&\frac{im\omega_{A_{\phi}}B_{\phi}^{(0)}}{\sqrt{\rho \mu_0}} =0,\nn\\
(\gamma + im \Omega  +\omega_{\eta}) B_R^{(0)}&-&i m\omega_{A_{\phi}} u_R^{(0)} \sqrt{\rho \mu_0} =0,\nn\\
(\gamma +im \Omega  +\omega_{\eta}) B_{\phi}^{(0)} &-&2\Omega {\rm Ro}B_R^{(0)}+2{\rm Rb}\omega_{A_{\phi}}\sqrt{\rho \mu_0}u_R^{(0)}\nn\\&-&i m\omega_{A_{\phi}} u_{\phi}^{(0)} \sqrt{\rho \mu_0} =0.
\ea

\subsection{Dimensionless parameters and dispersion relation}

The solvability condition for the system \rf{mri4ac} yields the dispersion relation
\be{dr1}
\det({\bf M}-\gamma {\bf E})=0,
\ee
where $\bf E$ is the $4\times4$ identity matrix and
\be{mri11}
{\bf M}=\left(
    \begin{array}{cccc}
      -im\Omega{-}\omega_{\nu} & 2\alpha^2\Omega & i  \frac{m \omega_{A_{\phi}}}{\sqrt{\rho \mu_0} } & -\frac{2\omega_{A_{\phi}}\alpha^2}{\sqrt{\rho \mu_0} } \\
      -2\Omega(1{+} {\rm Ro}) & -im\Omega{-}\omega_{\nu} & \frac{2\omega_{A_{\phi}}}{\sqrt{\rho \mu_0}}(1{+}{\rm Rb}) & i\frac{m \omega_{A_{\phi}} }{\sqrt{\rho \mu_0}} \\
      im \omega_{A_{\phi}}\sqrt{\rho \mu_0}  & 0 & -im\Omega{-}\omega_{\eta} & 0 \\
      -2{\omega_{A_{\phi}}}{\rm Rb}{\sqrt{\rho \mu_0} } & im \omega_{A_{\phi}}\sqrt{\rho \mu_0} & 2\Omega {\rm Ro} & -im\Omega{-}\omega_{\eta} \\
    \end{array}
  \right).
\ee
The polynomial \rf{dr1} has the same roots as the equation
\be{drdim}
\det({\bf M}{\bf T}-\gamma {\bf E}{\bf T})=0,
\ee
with ${\bf T}=\diag(1,1,(\rho \mu_0)^{-1/2},(\rho \mu_0)^{-1/2})$.

Now we introduce, in addition to the hydrodynamic $({\rm Ro})$ and
magnetic $({\rm Rb})$ Rossby numbers, the magnetic Prandtl
number $({\rm Pm})$,  the Reynolds $({\rm Re})$ and Hartmann
$({\rm Ha})$ numbers, as well as the modified azimuthal
wavenumber $n$:
\be{d3}
{\rm Pm}=\frac{\omega_{\nu}}{\omega_{\eta}},\quad
{\rm Re}=\alpha\frac{\Omega}{\omega_{\nu}},\quad
{{\rm Ha}}=\alpha\frac{\omega_{A_{\phi}}}{\sqrt{\omega_{\nu}\omega_{\eta}}},\quad
n=\frac{m}{\alpha}.
\ee
Dividing equation \rf{drdim} by ${\rm Re}$, we obtain
\be{d4}
p(\lambda)=\det\left({\bf H}-\lambda {\bf E}\right)=0,
\ee
with $\lambda=\gamma(\alpha \Omega)^{-1} $ and
\be{d5}
{\bf H}=\left(
  \begin{array}{cccc}
    -i n  -\frac{1}{\rm Re} & 2\alpha   & \frac{i n{\rm Ha}}{\sqrt{\rm Re Rm}} & \frac{-2\alpha{\rm Ha}}{\sqrt{\rm Re Rm}} \\
    -\frac{2(1+{\rm Ro})}{\alpha} & -i n  -\frac{1}{\rm Re} & \frac{2{\rm Ha}(1+{\rm Rb})}{\alpha\sqrt{{\rm Re Rm}}} & \frac{i n{\rm Ha}}{\sqrt{{\rm Re Rm}}} \\
    \frac{i n{\rm Ha}}{\sqrt{\rm Re Rm}} & 0 & -i n  -\frac{1}{{\rm Rm}} & 0 \\
    \frac{-2 {\rm Ha}{\rm Rb}}{\alpha \sqrt{\rm Re Rm}} & \frac{i n{\rm Ha}}{\sqrt{\rm Re Rm}} & \frac{2{\rm Ro}}{\alpha} & -i n  -\frac{1}{{\rm Rm}}\\
  \end{array}
\right),
\ee
where ${\rm Rm}={\rm Re}{\rm Pm}$ is the magnetic Reynolds number.

\subsection{Stability of the Chandrasekhar equipartition solution}
Observe that letting
\be{ces}
{\rm Rb} ={\rm Ro},\quad \Omega=\omega_{A_{\phi}}
\ee
in the equation \rf{mri11} and assuming
that $\omega_{\nu}=0$ and $\omega_{\eta}=0$ we find
that the dispersion relation \rf{dr1} has the following roots
$$
\gamma_{1,2}=0,\quad \gamma_{3,4}=-2i\omega_{A_{\phi}}(m\pm \alpha),
$$
which indicates marginal stability.

On the other hand note that ${\rm Ha}/\sqrt{\rm Re}$ in the matrix \rf{d5} is nothing
else but the square root of the interaction parameter (or Elsasser number)
\be{en}
{\rm N}:=\frac{{\rm Ha}^2}{\rm Re}.
\ee
Then, the condition $\Omega=\omega_{A_{\phi}}$, which is equivalent to
\be{hare}
{\rm Ha}=\sqrt{{\rm Re}{\rm Rm}},
\ee
is simply
\be{nrm}
{\rm N}={\rm Rm}.
\ee
With \rf{hare}, ${\rm Rb} ={\rm Ro}$, and ${\rm Re}\rightarrow \infty$ and ${\rm Rm}\rightarrow \infty$ the dispersion relation \rf{d4} yields
stable solutions in terms of $\lambda$
$$
\lambda_{1,2}=0,\quad \lambda_{3,4}=-2i(n\pm1).
$$

This should not
be surprising in view of the fact that at ${\rm Rb} ={\rm Ro}=-1$ \rf{ces}
is nothing else but the well-known Chandrasekhar equipartition
solution for the inviscid fluid of infinite electrical
conductivity with constant total pressure,
see (Chandrasekhar 1956, Chandrasekhar 1961).

Indeed, differentiating the constant total pressure condition $P=const$, we obtain
$$
\frac{\partial p_0}{\partial R}=-\frac{B_{\phi}^0}{\mu_0}\frac{\partial B_{\phi}^0}{\partial R}.
$$
Substituting this into \rf{m4d} yields
$$
R\Omega^2=-\frac{B_{\phi}^0}{\rho\mu_0}\frac{\partial B_{\phi}^0}{\partial R},
$$
which after letting $\Omega=\omega_{A_{\phi}}$ transforms into
$$
B_{\phi}^0\left(B_{\phi}^0+R\frac{\partial B_{\phi}^0}{\partial R}\right)=0.
$$
Taking into account that
$$
{\rm Rb}=\frac{1}{2B_{\phi}^0}\left(R\frac{\partial B_{\phi}^0}{\partial R}-B_{\phi}^0\right),
$$
we finally get
$$
2(B_{\phi}^0)^2(1+{\rm Rb})=0.
$$
Hence, from the assumption that $\Omega=\omega_{A_{\phi}}$ and that the total pressure $P$ is constant, we deduce that ${\rm Rb}={\rm Ro}=-1$.
Therefore, the conditions $\Omega=\omega_{A_{\phi}}$ and ${\rm Rb}={\rm Ro}=-1$ define the Chandrasekhar equipartition solution, which is marginally stable (Chandrasekhar 1956), as we have just demonstrated explicitly.

\subsection{Michael's criterion and its dissipative extension}
By applying the Bilharz stability criterion (Bilharz 1944, Kirillov 2013) to the complex dispersion relation \rf{dr1} with the matrix \rf{mri11}, we find that for stability against axisymmetric $(m=0)$ perturbations it is necessary and sufficient that
\be{michd}
{\rm Ro}>-1+{\rm Rb}\frac{\omega_{A_{\phi}}^2}{\Omega^2}\frac{\omega_{\nu}}{\omega_{\eta}}-\frac{\omega_{\nu}^2}{4\alpha^2\Omega^2}.
\ee
Note that the criterion \rf{michd} contains the ratio of
viscous to resistive frequencies and by this reason the
transition to the ideal case is not straightforward:
one needs taking first $\omega_{\eta}=\omega_{\nu}$ and
then letting $\omega_{\nu}$ tend to zero. The result is
the ideal Michael's criterion
\be{michi}
{\rm Ro}>-1+{\rm Rb}\frac{\omega_{A_{\phi}}^2}{\Omega^2}
\ee
or, equivalently (Michael 1954, Chandrasekhar 1961),
\be{miche}
\frac{d}{d R}(\Omega^2 R^4)-\frac{R^4}{\rho \mu_0}\frac{d}{d R}\left( \frac{B_{\phi}^0}{R}\right)^2>0.
\ee
Putting $\omega_{\nu}=0$ in the equation \rf{michd} yields
\be{michil}
{\rm Ro}>-1,
\ee
which is Michael's criterion corresponding
to the inductionless $({\rm Pm}=0)$ and inviscid $({\rm Re}\rightarrow \infty)$ case. In the next section we demonstrate that in the inductionless limit the Michael's criterion can easily be extended so as to comprise both axisymmetric $(m=0)$ and non-axisymmetric $(m\ne0)$ perturbations.

\section{Growth rates of AMRI}

\subsection{Inductionless case $({\rm Pm}=0)$}

Taking the definition \rf{en} into account in the dispersion relation \rf{d4} and assuming further
that the magnetic Prandtl number vanishes,
we find the roots of the polynomial $p(\lambda)$ in the following closed form
\be{clof}
\lambda=-in+{\rm N}(2{\rm Rb}-n^2)-\frac{1}{\rm Re}\pm2\sqrt{({\rm Rb}^2{+}n^2){\rm N}^2{+}in({\rm Ro}{+}2){\rm N}{-}1{-}{\rm Ro}}.
\ee
At small $\rm N$ we can expand $\lambda$ into the Taylor series
\be{exclt}
\lambda=-i(n\pm 2\sqrt{1+{\rm Ro}})-\frac{1}{\rm Re}+\left(2{\rm Rb}-n^2\pm \frac{n({\rm Ro}+2)}{\sqrt{{\rm Ro}+1}}\right){\rm N}+O({\rm N}^2),
\ee
evidencing that at  ${\rm Ro}>-1$ it is an inertial wave that will be destabilized by the weak azimuthal magnetic field (${\rm N}\ll 1$), no matter how close is its radial profile to the current-free profile $B_{\phi}\propto R^{-1}$. Destabilization of an inertial wave is characteristic of the inductionless MRI (Kirillov and Stefani 2010, 2011) in contrast to the Tayler instability.

Let $\lambda_r$ and $\lambda_i$ denote the real and the imaginary part of the complex root, i.e. $\lambda=\lambda_r+i\lambda_i$.
In the inviscid limit when ${\rm Re}\rightarrow \infty$, the
equation \rf{clof} yields explicit expressions for the growth rates of the perturbation
\ba{gra1}
\lambda_r = {\rm {N}}(2{\rm Rb}-n^2)\pm
\sqrt{2\left[({\rm Rb}^2+n^2){\rm {N}}^2-{\rm Ro}-1+\sqrt{D}\right]},
\ea
where
$$
D={\rm {N}}^2n^2(({\rm Ro}{+}1)^2{+}1{+}{\rm {N}}^2(2{\rm Rb}^2{+}n^2)){+}({\rm {N}}^2{\rm Rb}^2{-}{\rm Ro}{-}1)^2.
$$
Putting $\lambda_r=0$ in Eq.~\rf{gra1}, we find the condition for marginal stability
\be{gra4}
{\rm N}=\pm\frac{2\sqrt{-(n^2-4{\rm Rb}-4)((n^2-2{\rm Rb})^2({\rm Ro}+1)-({\rm Ro}+2)^2n^2)}}{n(n^2-4{\rm Rb}-4)(n^2-2{\rm Rb})}
\ee
that can be interpreted as a boundary of the instability domain in the $ {\rm N}-n$ plane. The threshold \rf{gra4} extends the Michael's criterion to arbitrary $n$ in the case when ${\rm Pm}=0$ and ${\rm Re}\rightarrow \infty$.

    \begin{figure}
    \begin{center}
    \includegraphics[angle=0, width=0.99\textwidth]{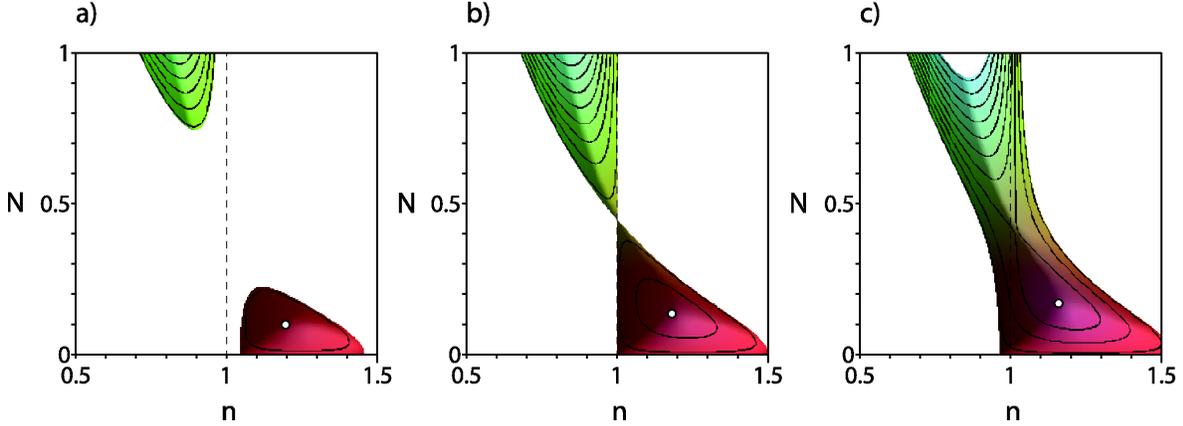}
    \end{center}
    \caption{The non-negative growth rates of AMRI  in the
    inductionless $(\rm Pm=0)$ and inviscid
    $({\rm Re} \rightarrow \infty)$ limit in
    projection onto the ${\rm N}-n$ plane at
    $\rm Ro=-0.75$ and (a) $\rm Rb=-0.76$, (b) $\rm Rb=-0.75$, and
    (c) $\rm Rb=-0.74$. The open circles mark the maximal growth rates
    (a) $\lambda_r\approx0.003$ at $n \approx 1.204$ and
    $ {\rm N} \approx 0.105 $, (b) $\lambda_r\approx0.005$
    at $n \approx 1.180$ and $ {\rm N} \approx 0.139 $,
    (c) $\lambda_r\approx0.008$ at
    $n \approx 1.151$ and ${\rm N} \approx 0.178 $.}
    \label{fig1}
    \end{figure}

In Fig.~\ref{fig1} the contour plots of the non-negative growth rates given by
Eq.~\rf{gra1} are shown in projection onto the ${\rm N}-n$ plane for the
special case of
Keplerian rotation, i.e. for ${\rm Ro}=-0.75$. The regions of the non-negative
growth rates are bounded by the curves \rf{gra4}. Note that at $n=0$ we have stability because ${\rm Ro}=-0.75>-1$ in accordance with the inductionless Michael's criterion \rf{michil} which follows also from \rf{gra4}. Observe that for ${\rm Rb}<-0.75$
there exist two instability regions, Fig.~\ref{fig1}(a), that touch each other
exactly at ${\rm Rb}=-0.75$,
Fig.~\ref{fig1}(b), and merge into one single region when ${\rm Rb}>-0.75$, Fig.~\ref{fig1}(c).
Remarkably, the intersection point visible in Fig.~\ref{fig1}(b) can be found
explicitly if we take ${\rm Ro}={\rm Rb}$ in Eq.~\rf{gra4}.
The marginal stability lines intersect then at the point $(n,{\rm N})$ with the coordinates
\be{gra5}
n=\pm 2 \sqrt{{\rm Rb}+1},\quad {\rm N}=\pm\frac{1}{2}\sqrt{\frac{-(3{\rm Rb}+2)}{({\rm Rb}+1)({\rm Rb}+2)}}.
\ee
For instance, the intersection point in Fig.~\ref{fig1}(b) has the coordinates
$n=1$, ${\rm N}=\frac{\sqrt{5}}{5}$. In view of the fact that $\alpha \in [0,1]$
and $n=m/\alpha$, where $m\ne0$ is an integer, the only physically meaningful
unstable region is situated in the domain $|n|\ge1$.

The growth rates reach their maxima inside the physically relevant
instability regions in Fig.~\ref{fig1}. The maximal growth rate
decreases from $\lambda_r\approx0.008$ at ${\rm Rb}=-0.74$
to $\lambda_r\approx0.003$ at ${\rm Rb}=-0.76$. With the
further decrease in $\rm Rb$ both the size of the lower instability
region and the maximal growth rate diminish until at
\be{gra6}
{\rm Rb}=-\frac{25}{32}=-0.78125
\ee
the lower instability region shrinks to a point which disappears at smaller
$\rm Rb$.

Indeed, the lower instability region disappears when the roots of
the equation \rf{gra4} become complex. Equivalently,
\be{gra7}
(n^2-2{\rm Rb})^2({\rm Ro}+1)-({\rm Ro}+2)^2n^2=0,
\ee
which factors out into the two equations quadratic in $n$,
\be{gra8}
n^2\pm\frac{{\rm Ro}+2}{\sqrt{{\rm Ro}+1}}n-2{\rm Rb}=0.
\ee
The roots $n$ of equations \rf{gra8} are complex if and only if their
discriminant is negative,
\be{gra9}
\frac{({\rm Ro}+2)^2}{{\rm Ro}+1}+8{\rm Rb}< 0,
\ee
which yields ${\rm Rb}<-\frac{25}{32}$ for the Keplerian
value of the Rossby number ${\rm Ro}=-\frac{3}{4}$.
On the other hand, the intersection of the marginal
stability curves \rf{gra5} exists at ${\rm N}\ne 0$ for
\be{gra6}
{\rm Ro}={\rm Rb}<-\frac{2}{3}.
\ee
At ${\rm Ro}={\rm Rb}
=-\frac{2}{3}$ the intersection occurs at ${\rm N}= 0$,
which means that, again, the lower instability region
disappears. This effect reminds one on the stability of
the Chandrasekhar equipartition solution for the
inviscid fluid of infinite electrical conductivity
with constant total pressure (Chandrasekhar 1956). Note,
however, that in contrast to (Chandrasekhar 1956) we
consider here a viscous and resistive fluid in the
inductionless limit that is not equivalent to the
ideal MHD case as one can see also on the example of
Michael's criterion discussed in Section 3.4.

\subsection{Instability condition}

    \begin{figure}
    \begin{center}
    \includegraphics[angle=0, width=0.99\textwidth]{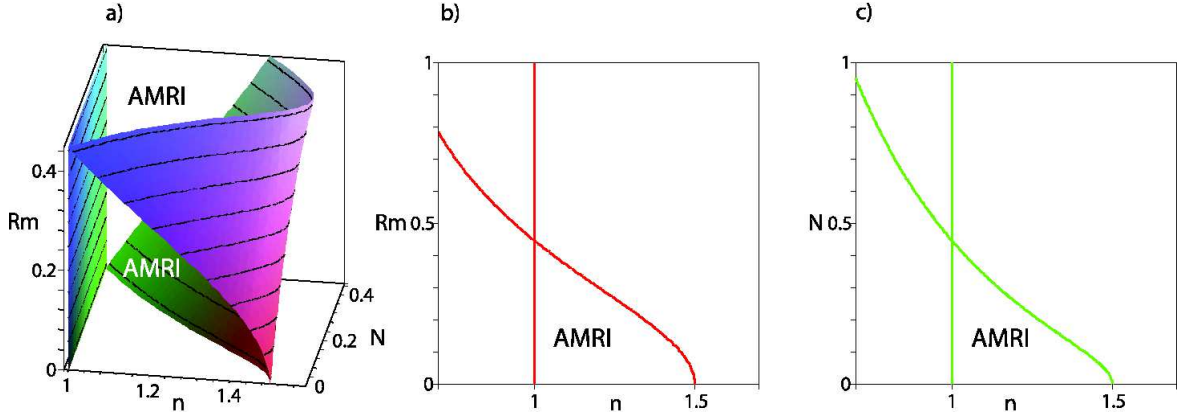}
    \end{center}
    \caption{Given ${\rm Re}\rightarrow \infty$ and ${\rm Rb}={\rm Ro}=-0.75$. (a) The domain of AMRI in the $(\rm n, N, Rm)$ space. Its cross-section (b) at $\rm N=0$ with the intersection at $\rm Rm =\sqrt{5}/5$ and (c) at $\rm Rm=0$ (inductionless case) with the crossing at $\rm N =\sqrt{5}/5$, cf. Fig.~\ref{fig1}(b).}
    \label{fig2}
    \end{figure}

Therefore, when ${\rm Ro}<-\frac{2}{3}$ and ${\rm Rb}<{\rm Ro}$, the instability domain in the ${\rm N}-n$ plane
consists of the two separate regions, see Fig.~\ref{fig1}(a). Since at
${\rm Rb}<-\frac{2}{3}$ we have $2 \sqrt{{\rm Rb}+1}>1$, then according to Eq.~\rf{gra5} the
lower instability region has physical meaning and thus corresponds to the azimuthal MRI.
This domain of AMRI exists, if
\be{gra9}
{\rm Rb}\ge-\frac{1}{8}\frac{({\rm Ro}+2)^2}{{\rm Ro}+1}.
\ee
Note that in the case when ${\rm Rb}=-1$, the condition of existence
of AMRI \rf{gra9} yields ${\rm Ro}\le2-2\sqrt{2}$, which corresponds to the well-known
Liu limit (Liu et al. 2006, Kirillov and Stefani 2010, Priede 2011, Kirillov et al. 2012).

Finally, we find the Taylor expansion of the growth rates near ${\rm N}=0$:
\ba{gra3}
\lambda_r &=&
\left(2{\rm Rb}-n^2\pm\frac{n({\rm Ro}+2)}{\sqrt{{\rm Ro}+1}}\right){\rm N}\nn\\&+&\frac{n({\rm Ro}+2)(4{\rm Rb}^2({\rm Ro}+1)-n^2{\rm Ro}^2)}{8({\rm Ro}+1)^{5/2}}{\rm N}^3+o({\rm N}^3).
\ea
Note that the coefficient in the  term that is linear in ${\rm N}$ is precisely the left
hand side of Eq.~\rf{gra8},
which determines the range of unstable values of $n$ at ${\rm N}=0$. On the other hand
the expansions \rf{gra3} demonstrate that
the hydrodynamically stable shear flow can be destabilized by an arbitrary small
azimuthal magnetic field, which happens in particular for
Keplerian flows, if ${\rm Rb}>-\frac{25}{32}$.

\subsection{Finite magnetic Reynolds number}
Let us replace in the matrix \rf{d5} the ratio ${\rm Ha}/\sqrt{\rm Re}$ with the $\sqrt{\rm N}$ according to the definition \rf{en} and then let
${\rm Re}\rightarrow \infty$. In the resulting dispersion relation we assume ${\rm Ro}={\rm Rb}$ and apply to it the Bilharz stability criterion (Bilharz, 1944, Kirillov, 2013). For ${\rm Ro}={\rm Rb}=-0.75$, the Bilharz criterion gives the domain of Azimuthal MRI presented in Fig.~\ref{fig2}(a). The domain is comprised between the plane $n= 2 \sqrt{{\rm Rb}+1}$ and the surface that intersects the plane along the two straight lines:
\be{gra5a}
n= 2 \sqrt{{\rm Rb}+1},\quad {\rm N}={\rm Rm}\pm\frac{1}{2}\sqrt{\frac{-(3{\rm Rb}+2)}{({\rm Rb}+1)({\rm Rb}+2)}}.
\ee
At $\rm N=0$ and ${\rm Rb}=-0.75$ the intersection point is at ${\rm Rm}=\sqrt{5}/5$, see Fig.~\ref{fig2}(b). On the other hand, the intersection point is at ${\rm N}=\sqrt{5}/5$ when ${\rm Rm}=0$ and ${\rm Rb}=-0.75$, as Fig.~\ref{fig2}(c) demonstrates. Naturally, the cross-section of the domain shown in Fig.~\ref{fig2}(c) exactly coincides with the domain of the inductionless AMRI shown in Fig.~\ref{fig1}(b).

Therefore, we have demonstrated that AMRI of Keplerian flows exists not only in the inductionless limit when $\rm Pm=0$ or, equivalently, $\rm Rm=0$, but also at finite values of the magnetic Reynolds number.

\section{Conclusions}

Using a WKB approach, we have considered the stability condition of
a rotating flow under the influence of an azimuthal magnetic field with
arbitrary radial dependence. Focusing on the case of small
magnetic Prandtl number, we have shown that Keplerian profiles
can be destabilized if only the azimuthal
field is shallow enough.
The necessary induction of $B_{\phi}(R)$ is comparably
small (${\rm Rb} \ge -0.78125$) so that the effect is definitely more
on the side
of the (inductionless) MRI than on the side of
the current-driven
Tayler instability with ${\rm Rb}=0$ (Seilmayer et al. 2012).

We have also shown that the point where
the hydrodynamic and the magnetic Rossby number are equal plays an
essential point for the connectedness of the instability domain.
With view on astrophysical applications one has to notice that
the shallower than $1/R$ profile of $B_{\phi}$ would need some
finite magnetic Reynolds number, while the Lundquist number can still
be arbitrarily small. Yet, the growth rate would then be
rather small, since it is proportional to the interaction parameter.
The consequences of our findings for those parts of accretion disks
with small magnetic Prandtl numbers are still to be elaborated.

Our results give strong impetus on dedicated MRI experiments
in which the magnetic Rossby number can be adjusted by using two
independent electrical currents, one through an central, insulated rod,
the second one through the liquid metal. A liquid sodium experiment
with such a possibility
is foreseen in the framework of the DRESDYN
project (Stefani et al 2012).

\section*{Acknowledgment}
This work was supported by Helmholtz-Gemeinschaft Deutscher
Forschungszentren (HGF) in frame of the Helmholtz Alliance LIMTECH,
as well as by Deutsche
Forschungsgemeinschaft in frame of the SPP 1488 (PlanetMag).

\section*{References}
\begin{harvard}

\item[]
Armitage P~J 2011 Dynamics of protoplanetary disks
{\it Ann. Rev. Astron. Astrophys.\/} {\bf 49} 195-236.

\item[]
Balbus S~A 2003 Enhanced angular momentum transport in accretion disks
{\it Ann. Rev. Astron. Astrophys.\/} {\bf 41} 555-597.

\item[]
Balbus S~A and Henri P 2008 On the magnetic Prandtl number behavior of accretion disks
{\it Astrophys. J.\/} {\bf 674} 408-414.

\item[]
Bilharz H 1944
Bemerkung zu einem Satze von Hurwitz
{\it Zeitschrift f\"ur Angewandte Mathematik und Mechanik} \textbf{24} 77-82.

\item[]
Chandrasekhar S 1956 On the stability of the simplest solution of the equations of hydromagnetics
{\it Proc. Nat. Acad. Sci. \/} {\bf 42} 273-276.

\item[]
Chandrasekhar S 1960 The stability of non-dissipative Couette flow in hydromagnetics
{\it Proc. Nat. Acad. Sci. \/} {\bf 46} 137-141.

\item[]
Chandrasekhar S 1961
Hydrodynamic and hydromagnetic stability, Oxford
University Press, Oxford, UK.

\item[]
Colgate SA et al 2011
High magnetic shear gain in a
liquid sodium stable Couette flow experiment:
A prelude to an $\alpha-\Omega$ Dynamo
{\it Phys. Rev. Lett.} {\bf 106}, 175003.

\item[]
Dobrokhotov S and Shafarevich A 1992
Parametrix and the asymptotics of localized solutions of the Navier-Stokes equations in $R^3$, linearized on a smooth
flow {\it Math. Notes\/} {\bf 51}, 47-54.

\item[]
Eckhardt B and Yao D 1995 Local stability analysis along Lagrangian paths
{\it Chaos, Solitons and Fractals\/} {\bf 5}(11), 2073-2088.

\item[]
Friedlander S and Vishik, M~M 1995
On stability and instability criteria for magnetohydrodynamics
{\it Chaos} {\bf 5}, 416-423.

\item[]
Herault J, Rincon F, Cossu C, Lesur G, Ogilvie G~I,
Longaretti P~Y 2011
Periodic magnetorotational dynamo action as a prototype
of nonlinear magnetic-field generation in shear flows
{\it Phys. Rev. E} {\bf 84}, 036321.

\item[]
Hollerbach R and R\"udiger G 2005
New type of magnetorotational instability in cylindrical Taylor-Couette flow
{\it Phys. Rev. Lett.} {\bf 95}(12), 124501.

\item[]
Hollerbach R, Teeluck V and R\"udiger G 2010
Nonaxisymmetric magnetorotational instabilities in cylindrical Taylor-Couette flow
{\it Phys. Rev. Lett.} {\bf 104}, 044502.

\item[]
Kirillov O N and Stefani F 2010
On the relation of standard and helical magnetorotational instability
{\it Astrophys. J.} {\bf 712}, 52.

\item[]
Kirillov, O N and Stefani  F 2011
Paradoxes of magnetorotational instability and their geometrical resolution.
{\it Phys. Rev. E} {\bf 84}(3), 036304.

\item[]
Kirillov O~N, Stefani  F and  Fukumoto Y 2012
A unifying picture of helical and azimuthal MRI, and the universal significance of the Liu limit
{\it Astrophys. J.} {\bf 756}, 83.

\item[]
Kirillov O N and  Stefani F 2013
Extending the range of the inductionless magnetorotational instability
{\it Phys. Rev. Lett.}, {\bf 111}, 061103 (5pp).

\item[]
Kirillov O N 2013 Nonconservative stability problems of modern physics, De Gruyter Studies in Mathematical Physics 14, De Gruyter, Berlin, Boston.

\item[]
Krueger E~R,  Gross A and Di Prima R~C 1966
On relative importance of Taylor-vortex and non-axisymmetric modes in flow between rotating cylinders
{\it J. Fluid Mech.\/} {\bf 24}(3), 521-538.

\item[]
Landman M~J and Saffman P~G 1987
The three-dimensional instability of strained vortices in a viscous fluid
{\it Phys. Fluids} {\bf 30}, 2339-2342.

\item[]
Liu W, Goodman J, Herron I and Ji H 2006
Helical magnetorotational instability in magnetized Taylor-Couette flow
{\it Phys. Rev. E} {\bf 74}(5), 056302.

\item[]
Michael, D~H 1954
The stability of an incompressible
electrically conducting fluid rotating about an axis when
current flows parallel to the axis {\it Mathematika}, {\bf 1}, 45-50.

\item[]
Petitdemange L, Dormy E, Balbus S~A 2008 Magnetostrophic
MRI in the earth's outer core
{\it Geophys. Res. Lett.} {\bf 35} L15305.

\item[]
Priede J 2011 Inviscid helical magnetorotational instability in cylindrical Taylor-Couette flow
{\it Phys. Rev. E\/} {\bf 84}, 066314.

\item[]
R\"udiger G, Gellert M, Schultz M and Hollerbach R 2010
Dissipative Taylor-Couette flows under the influence of helical magnetic fields
{\it  Phys Rev E} {\bf 82}, 016319.

\item[] R\"udiger G, Gellert M, Schultz M, Hollerbach R, Stefani F 2013
The azimuthal magnetorotational instability (AMRI) {\it Mon. Not. R. Astron. Soc.} submitted;
arXiv:1303.4621.

\item[]
Seilmayer M, Stefani F, Gundrum T, Weier T, Gerbeth G, Gellert M, R\"udiger G 2012
Experimental evidence for a transient Tayler instability in a cylindrical liquid metal column
{\it Phys. Rev. Lett.\/} {\bf 108} 244501.

\item[]
Stefani F, Gundrum T, Gerbeth G, R\"udiger G, Schultz M, Szklarski J, Hollerbach R 2006
Experimental evidence for magnetorotational instability in a Taylor-Couette flow under the
influence of a helical magnetic field
{\it Phys. Rev. Lett. \/} {\bf 97} 184502.

\item[]
Stefani F, Gerbeth G, Gundrum T, Hollerbach R, Priede J, R\"udiger G, Szklarski J 2009
Helical magnetorotational instability in a Taylor-Couette flow with strongly reduced Ekman pumping
{\it Phys. Rev. E\/}  {\bf 80}, 066303.

\item[]
Stefani F, Eckert S, Gerbeth G, Giesecke A, Gundrum T, Steglich C, Weier T, Wustmann B 2012
DRESDYN - A new facility for MHD experiments with liquid sodium
{\it Magnetohydrodynamics\/} {\bf 48} 103-113.

\item[]
Velikhov, E~P 1959
Stability of an ideally conducting liquid flowing between
cylinders rotating in a magnetic field {\it Sov. Phys. JETP - USSR\/} {\bf 9} 995-998.

\end{harvard}

\end{document}